\documentclass[aps,twocolumn,showpacs,floatfix]{revtex4}

\usepackage{dcolumn}
\usepackage{graphicx}
\usepackage{amsfonts}
\usepackage{amsmath,bm,amssymb}
\usepackage{comment}
\usepackage[usenames]{color}

\begin{document}

\title{Charge transfer, confinement, and ferromagnetism in
  LaMnO$_3$/LaNiO$_3$ (001)-superlattice}

\author{Alex Taekyung Lee}
\affiliation{Department of Physics, Korea Advanced Institute of Science and Technology, Daejeon 305-701, Korea }

\author{Myung Joon Han} \email{mj.han@kaist.ac.kr}
\affiliation{Department of Physics, Korea Advanced Institute of Science and Technology, Daejeon 305-701, Korea }
\affiliation{KAIST Institute for the NanoCentury, KAIST, Daejeon 305-701, Korea}

\date{\today }

\begin{abstract}
Using first-principles density functional theory calculations, we
investigated the electronic structure and magnetic properties of
(LaMnO$_3$)$_m$/(LaNiO$_3$)$_n$ superlattices stacked along
(001)-direction.  The electrons are transferred from Mn to Ni, and the
magnetic moments are induced at Ni sites that are paramagnetic in
bulk and other types of superlattices.  The size of induced moment is
linearly proportional to the amount of transferred electrons, but it
is larger than the net charge transfer because the spin and orbital
directions play important roles and complicate the transfer process.
The charge transfer and magnetic properties of the ($m$,$n$) superlattice
can be controlled by changing the $m/n$ ratio. Considering the
ferromagnetic couplings between Mn and Ni spins and the charge
transfer characteristic, we propose the (2,1) superlattice as the
largest moment superlattice carrying $\sim$8$\mu_B$ per formula unit.
\end{abstract}


\maketitle

{\it Introduction---} Recent advances in the layer-by-layer growth
technique of transition metal oxide (TMO) heterostructures have
created considerable research interest \cite{MRS, Hwang_review}.  In
TMO, multiple degrees of freedom ({\it i.e.}, charge, spin, orbital)
are coupled to each other, often creating novel material
characteristics such as high-temperature superconductivity and colossal
magneto resistance \cite{MIT-RMP}.  By making artificial
heterostructures of TMO, one can control those degrees of freedom and
band structures, and therefore create or design the new `correlated
electron' properties.  Previous TMO superlattice studies
\cite{Ohtomo-1,Okamoto,Ohtomo-2,Nakagawa,OrbReconst} have shown that
various unexpected material phenomena can be realized at the TMO
heterointerface such as magnetism
\cite{Brinkman,mag-LAOSTO-1,mag-LAOSTO-2} and superconductivity
\cite{Reyren}.

In this context, the superlattices composed of LaNiO$_3$ (LNO) and
LaMnO$_3$ (LMO) are of particular interest. A recent experiment by
Gibert and co-workers reported the exchange bias in LMO/LNO stacked
along the (111)-direction \cite{Gibert}.  Mn-to-Ni charge transfer is
expected at the LMO/LNO interface, which may cause a sizable magnetic
moment in the Ni ions even if LNO is paramagnetic in bulk
\cite{MIT-RMP} and many other superlattices
\cite{MJHan-opol,MJHan-LNOSTO,MJHan-LDAU}.  Importantly, however, it
is quite unclear if the same mechanism is also working in the LMO/LNO
superlattice stacked along (001).  Although the main concern of the
paper by Gibert et al. is the (111)-structure, their data does not
seem to support the same physics taking place in the (001)-stacked
LMO/LNO. On the other hand, a recent extensive experimental study by
Hoffman et al. \cite{Hoffman} reports that the same type of charge
transfer also occurs in the (001)-case, and the magnetic signals were
clearly observed from Ni sites. Furthermore, a recent theoretical work
by Dong and Dagotto \cite{Dong} suggests a different mechanism for the
induced magnetic moment ($M$) at Ni. Their tight-binding calculations
show that the induced magnetic moment in the (111)-superlattice is
better understood by the quantum confinement effect rather than by the
charge transfer. While the effect of confinement is strongest in the
(111)- and weakest in the (001)-structure, this study also raises an
important question regarding the induced Ni moment in the
(001)-superlattice.  However, due to the lack of first-principles
calculations for the (001)-structure, the detailed understanding of
this system has not yet been achieved.

In this paper, we examine the (001)-superlattice with first-principles
density functional theory calculations. Our calculations of
(LMO)$_m$/(LNO)$_n$ with several combinations of ($m$,$n$) clearly
show that the significant charge transfer occurs and the magnetic
moments are induced at Ni sites as in the (111)-case. Furthermore, the
size of an induced moment is linearly proportional to the amount of
charge transfer. However, the simple count of net electron transfers
cannot explain the size of the moment because the transfer process 
occurs in a complicated way that depends on spin and orbital
directions. The majority spin and $d_{3z^2-r^2}$ orbital are the major
channels in the electron transfer while the occupation changes in the
minority spin and $d_{x^2-y^2}$ orbital are not negligible.  We also
found strong ferromagnetic (FM) couplings between Ni and Mn,
whereas the Mn-Mn and Ni-Ni spins are antiferromagnetically aligned in
some cases. As a result, superlattices with ($m$,$n$)=(1,1), (1,2),
and (2,1) are always FM regardless of $U$ value, which can have 
important implications for application. Considering the amount of
charge transfers and FM couplings across the interface, the
(2,1) structure is proposed to have a largest moment of $\sim 8\mu_B$.

{\it Computation Details---}
For calculating (LMO)$_{m}$(LNO)$_{n}$ (001)-superlattices ($2\leqq m
\leqq3$, $2\leqq n \leqq4$), we used the projector augmented wave
(PAW) potentials \cite{PAW} and generalized gradient approximation
(GGA) proposed by Perdew \cite{PBE} for the exchange-correlation
functional, as implemented in the VASP code \cite{VASP}.  To study the
effect of electron correlation, we also used the GGA+$U$ scheme within the
rotationally invariant formalism and the double-counting formula, as
firstly proposed by Liechtenstein et al. \cite{LDA+U1}.  We used four
different sets of $U$ and $J$ for La-$4f$, Ni-$3d$, and Mn-$3d$
states: i) $U_{\textmd{all}}$ = $J_{\textmd{all}}$ = 0 ii)
$U_{\textmd{La}}$=6eV,$J_{\textmd{La}}$=0.5eV, $U_{\textmd{Ni}}$=6eV,
$J_{\textmd{Ni}}$=0.5eV, $U_{\textmd{Mn}}$=5eV,
$J_{\textmd{Mn}}$=0.5eV, iii) $U_{\textmd{La}}$=3eV,
$J_{\textmd{La}}$=0.5eV, $U_{\textmd{Ni}}$=3eV,
$J_{\textmd{Ni}}$=0.5eV, $U_{\textmd{Mn}}$=2.5eV,
$J_{\textmd{Mn}}$=0.5eV, and iv) $U_{\textmd{La}}$=0eV,
$J_{\textmd{La}}$=0eV, $U_{\textmd{Ni}}$=1eV, $J_{\textmd{Ni}}$=0eV,
$U_{\textmd{Mn}}$=4eV, $J_{\textmd{Mn}}$=1eV \cite{supplement}.  Note
that the last setup for $U$ and $J$ is the one used by Gibert et
al. for the (111)-structure \cite{Gibert}. While we are mainly
presenting and discussing the results from $U_{\textmd{Ni}}$=0 and 3
eV, it was found that the main claims and conclusions are not changed
in the other sets of parameters.  The wave functions were expanded in
plane waves with a kinetic energy cutoff of 500 eV.  We used a
\textbf{k}-point set generated by the 8$\times$8$\times$4
Monkhorst-Pack mesh for the (1,1) superlattice and used equivalent
\textbf{k}-points for other $(m,n)$-superlattices.  Ionic coordinates
were optimized until the residual forces were less than 0.01
eV/\AA. Wigner-Seitz radii of 1.286 and 1.323 \AA~ were used for
the projection of Ni and Mn atoms, respectively, as implemented in the
VASP-PAW pseudopotential. We assumed that the LMO/LNO superlattice is
grown on the SrTiO$_3$ substrate by setting the in-plane lattice
constant fixed at $a$=$b$=3.905\AA. We used the tetragonal supercell
and the optimized $c$-lattice parameters for each ($m$,$n$)
superlattice within the FM spin configuration.

\begin{figure}[t]
\begin{center}
\includegraphics[width=8cm, angle=0]{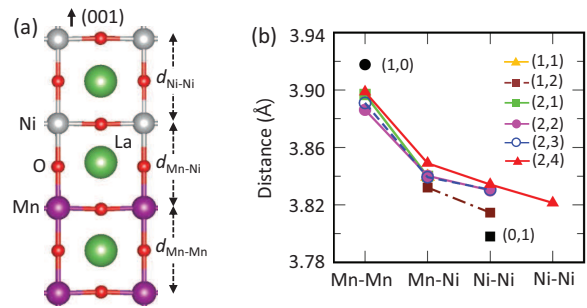}
\caption{ (Color online) (a) Atomic structure of (LMO)$_2$/(LNO)$_2$
  superlattice.  Grey, purple, green, and red colors stand for Ni, Mn, La,
  and O atoms, respectively.  $d_{\textmd{TM1-TM2}}$ denotes the
  distance between the TM1-plane and TM2-plane, where TM1 and TM2 are Mn
  or Ni atoms.  (b) The calculated $d_{\textmd{Mn-Mn}}$,
  $d_{\textmd{Mn-Ni}}$, and $d_{\textmd{Ni-Ni}}$ distances. The values
  for bulk LNO and LMO are indicated by ($m$,$n$)=(0,1) and (1,0),
  respectively. Note that in (2,4), two different types of
  $d_{\textmd{Ni-Ni}}$ exist. The shorter $d_{\textmd{Ni-Ni}}$ corresponds
  to the Ni-Ni distance between the two inner-most layers. }
\label{structure}
\end{center}
\end{figure}

%\section{Results and Discussion}
{\it Bulk and structural property---} The bulk LNO is known to have
low-spin $d^7$ electronic configuration and to remain as a
paramagnetic (PM) metal down to low temperature \cite{MIT-RMP}. The
local density approximation (LDA) and GGA calculation ($U$=0) predict
the correct PM ground state for the bulk phase and some other
superlattice structures such as LNO/LAO and LNO/STO \cite{MJHan-opol,
  MJHan-LNOSTO}, while LDA+$U$ predicts the local moment formation at
Ni site \cite{MJHan-LDAU,Pentcheva}. In our calculations, GGA+$U$
yields the Ni moment of 1.10 and 1.36 $\mu_B$ for $U$=3 and 6eV,
respectively.  In bulk LMO, Mn$^{3+}$ has high-spin $d^4$
configuration in which is, t$_{2g}^{\uparrow 3}$e$_g^{\uparrow 1}$. In
GGA ($U$=0) calculation, it is found that the small amount of
e$^{\uparrow}_g$ electron is transferred to t$^{\downarrow}_{2g}$ state
due to the down-spin t$_{2g}$ bands close to the Fermi level. The
calculated magnetic moment is increased from 3.46 $\mu_B$ at $U$=0 to
4.05$ \mu_B$ at $U$=6 eV.

The optimized out-of-plane lattice parameter of bulk LNO and bulk LMO
(with the fixed $a, b$ lattice of STO value) are found to be
$c_{\textmd{LNO}}=3.798$\AA~ and $c_{\textmd{LMO}}= 3.918$\AA,
respectively.  In the (LMO)$_m$/(LNO)$_n$ superlattice, the Ni-O-Ni
distance ($d_{\textmd{Ni-Ni}}$) and the Mn-O-Mn distance
($d_{\textmd{Mn-Mn}}$in Fig. 1(a)) along the $c$-axis are changed so that
$c_{\textmd{LNO}} < d_{\textmd{Ni-Ni}}$ and $c_{\textmd{LMO}} >
d_{\textmd{Mn-Mn}}$. As a result, the distances between the two
transition metals (TMs) in the superlattice are always larger than
$c_{\textmd{LNO}}$ and smaller than $c_{\textmd{LMO}}$, as clearly
shown in Fig.~\ref{structure}(b) for the $U$=0 case. It is noted that the
inner layer $d_{\textmd{Ni-Ni}}$ approaches to $c_{\textmd{LNO}}$ as
the thickness of LNO layers, $n$, increases.  We also found the same
trend in the $U>0$ results.

It is found that TM--O--TM bond angles in (LMO)$_m$/(LNO)$_n$ are 
generally not 180$^\circ$ \cite{supplement}.  The in-plane angle between
Mn-O-Mn ($\angle$Mn-O-Mn) in ($m$=1,$n$) superlattices is
$\approx$180$^\circ$ since these superlattices have mirror symmetry
with respect to the MnO$_2$-plane.  On the other hand, for
($m$=2,$n$) structure, $\angle$Mn-O-Mn decreases as $n$ increases.
Similarly, the in-plane angle between Ni--O--Ni ($\angle$Ni-O-Ni) at
the interface of ($m$=2,$n$) superlattices also decreases as $n$
increases, while $\angle$Ni-O-Ni$\approx$180$^\circ$ in (1,$n$)
superlattices. It is noted that $\angle$Ni-O-Ni is increased for the
bulk-like Ni atoms. To see the change of the out-of-plane TM--O--TM
bond angle and the possible octahedra rotations, we performed the
geometrical optimizations from distorted structures as starting
geometries in which the atomic positions are shifted toward the
in-plane oxygens (with no change in the lattice parameters). It was
found that the O atoms return to their original position and the
out-of-plane bond angles between Ni--O--Ni remain as 180$^\circ$.

\begin{figure}[t]
\begin{center}
\includegraphics[width=8cm, angle=0]{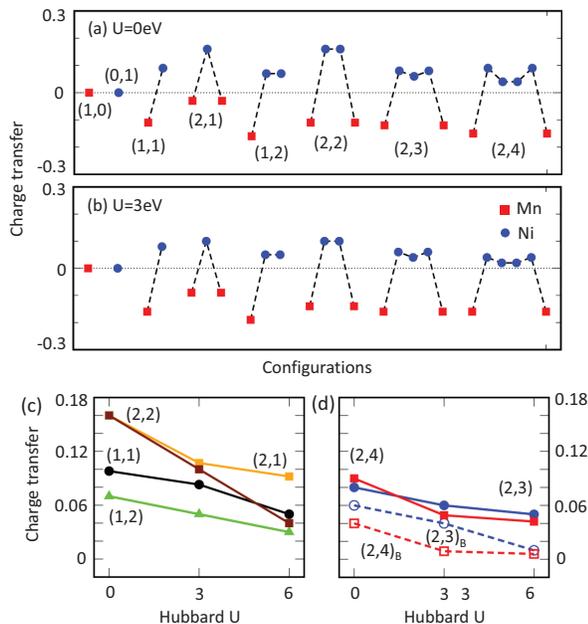}
\caption{ (Color online) The amount of charge transfer in
  (LMO)$_m$/(LNO)$_n$ superlattices (denoted by ($m$,$n$)) calculated
  with (a) $U$=0 and (b) $U$=3eV.  Red boxes and blue circles
  represent the calculated charges of Mn and Ni atoms, respectively.
  Zero-charge indicates the values from the bulk LMO and LNO.  (c)-(d)
  The amount of transferred charge to Ni as a function of $U$.
  (2,3)$_B$ and (2,4)$_B$ stand for the bulk-like Ni atom in the (2,3)
  and (2,4) superlattices, respectively, while others (with no
  subscript) refer to the interface Ni. }
\label{charge}
\end{center}
\end{figure}

{\it Charge transfer and Ni magnetic moment---}
Figs.~\ref{charge}(a)-(b) summarize the calculated result of the charge
transfer between Ni and Mn for several ($m$,$n$) combinations of
(LMO)$_m$/(LNO)$_n$, where up and down panels correspond to the gain
and loss of electrons, respectively. The number of TM-$d$ electrons in
the bulk LNO and the bulk LMO are set to be zero as a reference point
for Ni and Mn charge, respectively.  The results correspond to the most
stable spin configuration among all possible spin orders for given
($m$,$n$) structures.  A clear common feature is that the electrons
are transferred from Mn to Ni.  Although the amount of charge transfer
in the (111)-case is not given in Ref.~\onlinecite{Gibert}, we expect
that the charge transfer in the (111)-superlattice is larger than that
of the (001)-case because the (111)-interface creates more Mn--O--Ni bonds
than the (001)-interface does. This point is also reflected in the result
of the magnetic moment, which will be discussed further. The
transferred electrons mostly reside at the interface Ni sites and the
valence change in the bulk-like (inner layer) Ni is relatively small,
as clearly seen in the result of (2,3) and (2,4).

Note that, since Mn donates electrons to Ni, the amount of charge
transfer and the Ni valency can be controlled by changing the
superlattice composition ($m$,$n$). For a larger ratio of $m$/$n$,
the induced change in the Ni valence becomes larger while that for the
smaller $m$/$n$ the change smaller. By comparing (1,1) structure with
(2,1), one can find that the Ni-$d$ occupation is larger in (2,1),
where the two Mn ions can provide electrons to one Ni. The same
feature is confirmed by comparing (2,2) with (2,3).  As we will
discuss below, the transferred electrons induce the magnetic
moment at Ni, and therefore, the magnetism can also be controlled by
changing the superlattice compositions ($m$,$n$).

The main feature regarding the charge transfer is maintained even when the
on-site correlation $U$ is turned on as shown in Fig.~\ref{charge}(b).
The same curve shapes are found as in the $U$=0 results
(Fig.~\ref{charge}(a)), indicating the same type of charge transfer.
The effect of $U$ is to reduce the amount of charge transfer. $U$=6 eV
results \cite{supplement} are also found to be consistent with $U$=3
(Fig.~\ref{charge}(b)).  The effect of correlations that reduces the
charge transfer is more clearly seen in Figs.~\ref{charge}(c) and (d),
where the increase of Ni-$d$ occupation (with respect to the bulk
value) is plotted as a function of $U$.  The decreasing feature is
evident for all compositions of ($m$,$n$) and for both interfacial and
bulk-like Ni. It is noted that, in the bulk-like Ni sites, the valence
change caused by charge transfer is close to zero if $U$ is large
enough (see the dashed line in Fig.~\ref{charge}(d) at $U$=6 eV).

\begin{figure}[t]
\begin{center}
\includegraphics[width=8cm, angle=0]{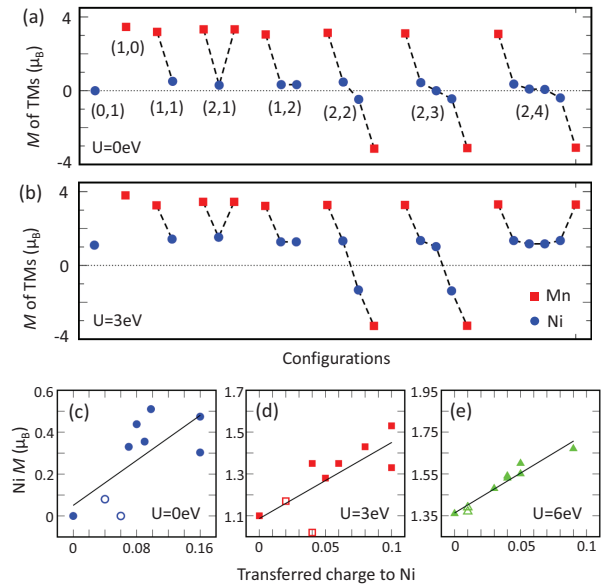}
\caption{ (Color online) (a)-(b) The calculated magnetic moments of Mn
  and Ni in the superlattices with (a) $U$=0 and (b) $U$=3eV.  Red
  boxes and blue circles represent the Mn and Ni values, respectively.
  (c)--(e) The calculated Ni magnetic moment as a function of
  transferred charge to Ni for (c) $U$=0eV, (d) $U$=3eV, and (e)
  $U$=6eV.  Filled symbols represent the interfacial atoms, whereas
  open symbols represent the bulk-like ones}
\label{spin}
\end{center}
\end{figure}

For the (111)-superlattices of LMO/LNO \cite{Gibert}, it is reported
that the magnetic moment is induced at Ni, which is originally
paramagnetic in bulk, and the exchange bias is manifested by this
induced moment.  For the (001)-case, however, it is unclear if the Ni
atoms are spin polarized.  It seems that two experimental studies
arrive at the different conclusions regarding this question for
the (001)-case.  Gibert et al.~\cite{Gibert} reported that the exchange
bias is not manifested in the case of the (001)-interface, while Hoffman et
al.~\cite{Hoffman} clearly observed the magnetic signals. Therefore, a
detailed theoretical analysis is required to understand the magnetic
property of the (001)-structures.

Our calculations clearly show that the Ni magnetic moment is also
induced in (001)-superlattices. After calculating all the possible
spin orders for given ($m$,$n$), we present the most stable spin
configurations in Figs.~\ref{spin}(a)-(b), where up and down panels
represent majority (up) and minority (down) spin, respectively. It is
noted that the Ni ions have non-zero spin moment even in the $U$=0
calculations (Fig.~\ref{spin}(a)).

The calculated Ni moment is $\sim$0.08--0.51 $\mu_B$ at $U$=0, and
enhanced up to $\sim$1.10--1.53 $\mu_B$ at $U$=3 eV.  It is noted
that, for the (2,2) superlattice, the calculated value of
$M_{\textmd{Ni}}$ is 0.47 $\mu_B$ at $U$=0, similar to the experiment
$\sim$0.35 $\mu_B$ \cite{Hoffman}.  As for $M_{\textmd{Mn}}$ , there
is a significant difference between the calculated value of
3.14$\mu_B$ and the experimental one ($\sim$2$\mu_B$ \cite{Hoffman}).
Although the origin of this discrepancy is unclear, we emphasize that
the calculated value is in good agreement with Mn charge status of 4+,
which is supported both by our calculation (see Fig.~\ref{charge}) and
the x-ray absorption spectroscopy data in Ref.~\onlinecite{Hoffman}
\cite{supplement}.  It should be noted that for bulk LNO and other
superlattices such as LNO/LAO \cite{MJHan-opol,MJHan-LDAU} and LNO/STO
\cite{MJHan-LNOSTO}, GGA (or LDA; $U$=0) predicts zero moment for Ni.
Therefore, our result of finite Ni moments at $U$=0 is a clear
evidence for the induced net moment.

It is instructive to compare the magnetic property of the (001)-superlattice 
with (111). In both cases, Ni magnetic moment is induced
and coupled to Mn spins ferromagnetically. Also, Mn-Mn and Ni-Ni
coupling is antiferromagnetic (AFM) in (2,2) structure. The notable
differences are found in the size of the magnetic moments. According to
Gibert et al. \cite{Gibert}, $1.1 \leq M_\textmd{Ni} \leq 1.4\mu_B$
for the (111)-case.  These values are much larger than our results for
the (001)-interface with $U$=0 (see Fig.~\ref{spin}(a)). To be more
precise, we performed the calculations with the same $U$ and $J$
values as used in Ref.~\onlinecite{Gibert}.  The result clearly shows
that the calculated Ni moment is always smaller in (001)-superlattice
than in (111) by $\sim 0.3\mu_B$, while the Mn moment is larger in
(001) by $\sim 0.3\mu_B$. Note that this trend is also consistent with
the charge transfer feature, as discussed above.

\begin{figure}[t]
\begin{center}
\includegraphics[width=8cm, angle=0]{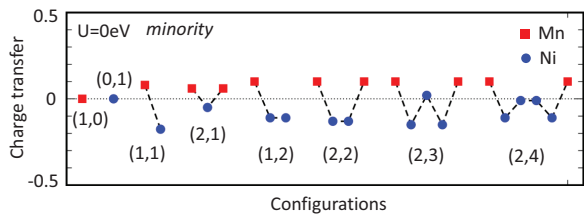}
\caption{ (Color online) The calculated charge transfer for the
  minority-spin ($U$=0eV).  Red boxes and blue circles represent the
  Mn and Ni values, respectively.  Note that the charge transfer shape
  is opposite of that of Fig. 2(a).  }
\label{renorm}
\end{center}
\end{figure}

The origin of the induced moment in LMO/LNO is under debate
\cite{Gibert,Hoffman,Dong}.  Gibert et al. speculated about the charge
transfer and two-dimensional confinement as a possible origin of the
induced Ni moment and seem to have concluded that neither of them
plays a significant role \cite{Gibert}.  On the other hand, in an
interesting recent study, Dong and Dagotto suggest that the induced
moment is better understood as a result of the spin-dependent quantum
confinement rather than the charge transfer especially for the case of
the (111)-superlattice.  This confinement effect is shown to be
strongest in the (111)-interface and weakest in (001).  Thus, further
study seems necessary for the (111)-case, and it is important to
understand the role of confinement and charge transfer in the
(001)-case.

To address this point, we present the induced Ni moment as a function
of the amount of electron transfer in Figs. \ref{spin}(c)-(e). A linear
dependency is quite clear, especially for the non-zero $U$ calculations,
and the $U$=0 result is not very far from the linear fit. This point
suggests that the charge transfer is the key origin of the induced Ni
moments in case of the (001)-superlattices. It is noted that while the
size of Ni moments is larger for $U>0$, the moment enhancement by
heterostructuring is larger in $U$=0 because the correlation $U$
reduces the charge transfer. It is also consistent with the picture of
charge-transfer-driven moment formation.

{\it Spin and orbital dependency---} It is interesting to note that
the charge transfer between Mn and Ni takes place in a different way
in the minority-spin bands from the majority.  It is found that the
number of minority-spin electrons at Mn is enhanced in the
superlattice (compared to the bulk value) whereas that of Ni is mostly
reduced, which is opposite to the case of majority-spin.
Fig.~\ref{renorm}(a) clearly shows that the sign of charge transfer is
reversed in the minority spin case (compare Fig.~\ref{renorm} with
Fig.~\ref{charge}(a)).
As a result, the induced Ni moment is larger than the \emph{net} charge transfer,
and we note that this effect is largest for the (1,1) superlattice
with the largest induced Ni moment.

Importantly, the total amount of charge transfer is dominated by
majority-spin channels while the change in the minority-spin electron
occupations is relatively small.  We note that this point is
consistent with the spin-dependent quantum confinement picture
suggested by Dong and Dagotto \cite{Dong}, even though their analysis
is best applied to the (111)-interface, and the effect is relatively
weak in the (001)-case.  In this picture, the majority-spin Ni-$e_g$
wavefunction is more widely spread out while the minority-spin
electron is localized. The delocalized feature of the majority-spin
bands and the more overlap with the neighboring up-spin Mn bands are
consistent with our results that the charge transfer occurs through
the majority-spin channel.  For the majority spins, the Ni-$d$
occupations decrease as $U$ increases.  In the minority spin bands,
the occupation changes are much smaller \cite{supplement}.  Although
the occupation changes are quite small, interestingly the amount of
occupation enhancement is found to increase as $U$ increases, which is
an opposite trend to the majority spin case.

By analysing the orbital occupations, we found that the major charge
transfer channel is $d_{3z^2-r^2}$ orbital. Compared to the bulk
value, the majority-spin Mn-$d_{3z^2-r^2}$ occupation is reduced by
$\sim$0.07--0.16 depending on ($m$,$n$) ($U$=0 eV).  The change in
Mn-$d_{x^2-y^2}$ occupations is less reduced as $\sim$
0.02--0.11. Due to its wavefunction shape, $d_{3z^2-r^2}$ can have
more overlaps and be a more efficient channel for this process.  The
same feature is found in the change of Ni-$e_g$
occupations. Ni-$d_{3z^2-r^2}$ occupations get enhanced by $\sim$
0.12--0.18 while the Ni-$d_{x^2-y^2}$ by $\sim$ 0.06--0.10. The main
features of this spin and orbital dependent charge transfer are
maintained also in the GGA+$U$ calculations \cite{supplement}.

{\it Designing ferromagnetic superlattices---} Making an FM TMO
superlattice with a large magnetic moment and a high Curie temperature
is an important issue for applications
\cite{Luders,Schuster,Dang,Dang2}.  It is noted in our system that the
induced Ni moment can ferromagnetically align with Mn spins as in the
(111)-superlattice of LMO/LNO \cite{Gibert}. As summarized in
Figs.~\ref{spin}(a)-(b), our calculations show that in the
(001)-superlattice FM coupling across the interface ({\it i.e.,}
between Ni and Mn) is always favored energetically. That is, the
interface FM spin arrangements always have less total energy than AFM
ones, regardless of the other parts of spin orders and independent of
$U$ values (see Figs.~\ref{spin}(a)-(b)).  For example, the FM (1,1)
superlattice has the lower total energy than the AFM one by 130
meV/(LNO)$_1$(LMO)$_1$ at $U$=3eV, which corresponds to the magnetic
coupling $J_{\textrm{Ni-Mn}}$= 58 meV with $S_{\textmd{Mn}}$=1.6 and
$S_{\textmd{Ni}}$=0.7.  While the Ni-Mn spins are always aligned
ferromagnetically, the Mn-Mn and Ni-Ni couplings are either AFM or FM
depending on $U$ and ($m$,$n$) (see Figs.~\ref{spin}(a)-(b))
\cite{supplement}.

Our result has an interesting implication in regard to the design of
the magnetism of superlattices. Since the interface (Ni-Mn) coupling is
always FM, the superlattice compositions of (1,1), (1,2), and (2,1)
should be FM carrying large total moments. Furthermore, (2,1) structure is
expected to have the largest moment, especially because the amount of
charge transfer will be largest in this case as we discussed already
(that is, largest $m$/$n$ ratio). Therefore, based on our analysis for
charge transfer and magnetic coupling, one can expect the
(2,1) structure to be the large moment FM superlattice.  The
calculated total moments of (1,1), (2,1), and (1,2) are 4.14, 7.71, and
4.14$\mu_B$, respectively, at $U$=0 eV.  It is noted that the
significant amount of magnetic moment can actually be controlled by
changing the ($m$,$n$) compositions.  The calculated moments by $U$=3
are 5.0, 9.0, and 6.0$\mu_B$ for (1,1), (2,1) and (1,2), respectively,
being consistent with $U$=0 results. The suggested guiding principle
based on the FM coupling and the charge transfer can be useful to
design the magnetic superlattices.

%\section{Summary}
{\it Summary ---} Our first-principles calculations show that the
magnetic moments are induced at Ni atoms in the (001)-oriented
(LMO)$_m$/(LNO)$_n$. The induced Ni moment is governed by the electron
transfer from Mn to Ni and the amount of charge transfer increases as
$m/n$ increases. Spin and orbital directions also play important role.
Our analysis based on the FM couplings between Mn and Ni and the
charge transfer features can provide an useful designing principle for
the magnetic TMO superlattices.

%\section{Acknowledgments}
We thank Heung-Sik Kim, Jason Hoffman and Anand Bhattacharya for
helpful discussion.  This work was supported by the National Institute
of Supercomputing and Networking / Korea Institue of Science and
Technology Information with supercomputing resources including
technical support (KSC-2013-C2-005).

\end{document}